\documentclass[preprint,showpacs,preprintnumbers,amsmath,amssymb]{revtex4}

\usepackage{graphicx}
\usepackage{dcolumn}
\usepackage{bm}
\usepackage{verbatim}
\usepackage{epsfig}
\usepackage{epstopdf}

\begin{document}


\title{A self-adjoint arrival time operator inspired by measurement models}

\author{J.J.Halliwell}%

\email{j.halliwell@imperial.ac.uk}

\author{J.Evaeus}

\author{J.London}

\author{Y.Malik}

\affiliation{Blackett Laboratory \\ Imperial College \\ London SW7
2BZ \\ UK }



\begin{abstract}
We introduce an arrival time operator which is self-adjoint and, unlike previously proposed arrival time operators, has a close link to simple measurement models. Its spectrum leads to an arrival time distribution which is a variant of the Kijowski distribution (a re-ordering of the current) in the large momentum regime but is proportional to the kinetic energy density in the small momentum regime, in agreement with measurement models.  A brief derivation of the latter distribution is given. We make some simple observations about the physical reasons for self-adjointness, or its absence, in both arrival time operators and the momentum operator on the half-line and we also compare our operator with the dwell time operator.
\end{abstract}


\maketitle



\newcommand\beq{\begin{equation}}
\newcommand\eeq{\end{equation}}
\newcommand\bea{\begin{eqnarray}}
\newcommand\eea{\end{eqnarray}}

\def\xb{{\bar x}}
\def\A{{\cal A}}
\def\D{\Delta}
\def\H{{\cal H}}
\def\E{{\cal E}}
\def\p{\partial}
\def\la{\langle}
\def\ra{\rangle}
\def\ria{\rightarrow}
\def\x{{\bf x}}
\def\y{{\bf y}}
\def\k{{\bf k}}
\def\q{{\bf q}}
\def\p{{\bf p}}
\def\P{{\bf P}}
\def\r{{\bf r}}
\def\s{{\sigma}}
\def\a{\alpha}
\def\b{\beta}
\def\e{\epsilon}
\def\U{\Upsilon}
\def\G{\Gamma}
\def\om{{\omega}}
\def\Tr{{\rm Tr}}
\def\ih{{ \frac {i} { \hbar} }}
\def\trho{{\rho}}
\def\au{{\underline \alpha}}
\def\bu{{\underline \beta}}
\def\pp{{\prime\prime}}
\def\id{{1 \!\! 1 }}
\def\half{\frac {1} {2}}
\def\jjh{j.halliwell@ic.ac.uk}

\section{Introduction}


The arrival time problem in quantum mechanics is the question of determining the probability that
an incoming wave packet, for a free particle, arrives at the origin in a given time interval \cite{Allcock,TQM,TQM2,rev1,rev2}. Classically, for a particle with initial position $x$ and momentum $p$, the arrival time is given by the quantity
\beq \label{classical}
\tau = -\frac{mx}{p}.
\eeq
The quantum problem is most simply solved using the spectrum of an operator corresponding to this quantity, such as that first studied by Aharonov and Bohm \cite{AB},
\beq \label{ABoperator}
\hat{T}_{AB} = -\frac{m}{2}\left(\hat{x}\frac{1}{\hat{p}} +\frac{1}{\hat{p}}\hat{x}\right).
\eeq
A heuristic result due to Pauli \cite{Pauli} (significantly updated by Galapon \cite{Gal,Gal2}) indicates that an object such as this, which is conjugate to a Hamiltonian with a semi-bounded spectrum, cannot be self-adjoint.
Indeed we find that its eigenstates, which in the momentum representation (with $\hat x \rightarrow i \hbar \partial / \partial p$) are given by
\beq
\phi_{\tau}(p) = \left(\frac{|p|}{2\pi m\hbar}\right)^{\half}e^{i\frac{p^{2}}{2m\hbar}\tau},
\label{ABstates}
\eeq
are complete but not orthogonal. There is a POVM associated with these states \cite{Gia} from which an arrival time
distribution can be constructed and it coincides with that postulated by Kijowski \cite{Kijowski}, namely
\bea
\Pi_{K}(\tau) &=& |\la \psi | \phi_\tau \ra|^{2} \nonumber \\ &=& \frac{1}{m} \la \psi_{\tau} | |\hat{p}|^{\half} \delta(\hat{x}) |\hat{p}|^{\half}| \psi_{\tau}\ra
\label{Kij}
\eea
(where $ \Pi (\tau) d \tau $ is the probability of arriving at the origin between $\tau $ and $ \tau + d \tau $), for which there is some experimental evidence \cite{Meas}.
This is related by a simple operator re-ordering to the quantum-mechanical current at the origin, $ \langle \hat J(t) \rangle $, the result expected on classical grounds, where the current operator is given by
\beq
\hat J(t) = \frac {1} {2m} \left( \hat p \delta ( \hat x(t) ) + \delta ( \hat x(t) ) \hat p \right),
\label{cur}
\eeq
with $ \hat x (t) = \hat x + \hat p t / m $. This picture, the standard one, is nicely summarized in Ref.\cite{stan} and some developments of it and explorations of the underlying mathematics are described in Refs.\cite{Gal,Gal2,Gal3,recent}.

The purpose of the present paper is to make two contributions to the standard picture presented above. The first is to discuss three simple self-adjoint modifications of the Aharonov-Bohm operator, discuss the relationship with the momentum operator on the half-line, and identify the underlying physically intuitive reasons why some of these operators are self-adjoint and some not.
The second and main result is to present a new self-adjoint arrival time operator which has a much closer link to models of measurement than any previously studied operators.

\section{Self-Adjoint Arrival Time Operators.}

The lack of self-adjointness of Eq.(\ref{ABoperator}) is not necessarily a problem but nevertheless a number of efforts have been made to restore it. Here, we take a simple approach and note that self-adjointness may be achieved by a number of simple modifications of the states Eq.(\ref{ABstates}).
The states
\beq \label{dmstates}
\phi_{\tau}(p) = \left(\frac{|p|}{2\pi m\hbar}\right)^{\half}e^{i\e(p)\frac{p^{2}}{2m\hbar}\tau},
\eeq
where $\epsilon(p)$ is the sign function, are orthogonal and complete and so are eigenstates of a self-adjoint operator. This operator, first considered by Kijowski \cite{Kijowski} and subsequently explored at length by
Delgado and Muga \cite{DM}, may be written
\beq
\hat{T}_{KDM}= -\frac{m}{2}\left( \hat{x}\frac{1}{|\hat{p}|} +\frac{1}{|\hat{p}|}\hat{x}\right),
\label{DMop}
\eeq
and is a quantization of the classical expression $ - m x / |p|$. It is also usefully written in the representation using the pseudo-energy, $ \xi =  p | p | / 2m $, where we have
\beq
\hat{T}_{KDM} = -i \hbar \frac { \partial } { \partial \xi}
\eeq
(which acts on states $ \Phi (\xi) = (m/p)^\half \phi (p) $), and is self-adjoint since $\xi $ takes an infinite range.
This is in contrast to the Aharonov-Bohm operator which, in the energy representation, has the form
\beq
\hat{T}_{AB} = \left( -i \hbar \frac { \partial} { \partial E} \right) \oplus  \left( -i \hbar \frac { \partial} { \partial E} \right)
\eeq
where the two parts of the direct sum refer to the positive and negative momentum sectors. Its lack of self-adjointness is due to the fact that $E>0$, as is frequently noted. (See for example, Ref.\cite{EgMu}).

A second modification of the Aharonov-Bohm operator is to superpose opposite values of
$\tau$ in Eq.(\ref{ABstates}) and then note that the subsequent states, which are proportional to $|p|^\half \sin ( p^2 \tau / 2 m \hbar ) $ are orthogonal and are the eigenstates of the self-adjoint operator
\beq
\hat T_{MI} = \sqrt{\hat{T}_{AB}^{2}},
\eeq
considered by de la Madrid and Isidro \cite{MI}. This is a quantization of $ m |x|/ |p|$.
A third modification is to note that the orthogonality of the states  $|p|^\half \sin ( p^2 \tau / 2 m \hbar ) $ is not affected by restriction to positive or negative momenta so we may consider these two sectors separately and as a consequence the operator
\beq
\hat T_3 = \theta (\hat p) \hat T_{MI} \theta (\hat p) -\theta (-\hat p) \hat T_{MI} \theta (-\hat p),
\eeq
is self-adjoint. This operator, which does not seem to have been noted previously,
is a quantization of the classical expression $ m |x|/ p$.

These three examples all side-step the Pauli theorem since they do not have canonical commutation with the Hamiltonian. Furthermore, they all give probability distributions which are simple variants on the Kijowski distribution, the expected result, as is easily deduced from their eigenstates.

From these three examples we make the following simple observation. The Aharonov-Bohm operator arises from the quantization of the classical expression $ - m x/p$ and is not self-adjoint. However, quantizing any of the three classical expressions $ - m x /|p|$, $ m |x|/|p|$ or $ m |x|/p$ leads to a self-adjoint operator. Hence,
self-adjoint modifications of the Aharonov-Bohm operator are easily obtained by relinquishing just one or two bits of information, namely the sign of $x$, or $p$, or the signs of both. The relinquished information is essentially the specification of whether the particle is incoming ($x$ and $p$ with opposite signs) or outgoing ($x$ and $p$ with the same sign). Of course in practice we are usually interested in the arrival time for a given state, for which this information is already known, at least semiclassically, so from this point of view the difference between the Aharonov-Bohm operator and its self-adjoint variants may not be important. Nevertheless it is of interest
to uncover the underlying origins of self-adjointness or its absence and the above properties give some useful clues.

Physically speaking, self-adjointness or its absence are about precision. A self-adjoint operator has orthonormal eigenstates and an associated projection operator onto a range of its spectrum. Projections onto different ranges have zero overlap.
An operator that is not self-adjoint has non-orthogonal eigenstates and has at best a POVM onto a range of its spectrum. Two POVMs localizing onto different ranges will have a small overlap which means there is intrinsic imprecision in the specification the ranges they localize onto.
To understand the origin of the lack of self-adjointness in the Aharonov-Bohm operator we would like to find a physically intuitive understanding of where this imprecision comes from.

\section{The Momentum Operator on the Half-Line}

To understand the above issue,
we turn to the frequently-studied situation of the momentum operator on the half-line $x>0$ \cite{Mom,half,ShHo}. There, the momentum operator cannot be made self-adjoint since it generates translations into negative $x$. However, $\hat p^2$ can be made self-adjoint, with suitable boundary conditions, and therefore, by the spectral theorem, $ | \hat p | $ can be made self-adjoint. Hence just by relinquishing information about the sign of $\hat p$ a self-adjoint operator is obtained.
The obstruction to self-adjointness on the half-line therefore lies in the sign function of $\hat p$. Differently put, the problem is that the operator $\theta (\hat x) \theta (\hat p) \theta (\hat x)$ cannot be made self-adjoint. For similar reasons, we also note that the position operator on the positive momentum sector cannot be self-adjoint. A POVM for the momentum operator on the half-line may be constructed, but this is not directly relevant to what we do here \cite{ShHo}.

We propose that there is a simple physical way of understanding the underlying imprecision linked to this lack of self-adjointness. Suppose we tried to measure the momentum. Let us therefore consider
a simple measurement model of momentum on the half-line  $x>0$ using sequential position measurements, from which information about momentum can be deduced. Similar approaches to calculating the time-of-flight momentum have been given elsewhere \cite{ToF} and we make use of these results, but adapted to the case of propagation in the region $x>0$.
We suppose we have an initial incoming state $ \psi $ at time $t_0$ consisting of spatially very broad gaussian, close to a plane wave, of momentum $p_0<0$
and we ask if it passes through a spatial region $ [\xb_1 - \Delta, \xb_1 + \Delta]$ in $x>0$ at time $t_1$ and at a later time $t_2$ through a spatial region $[\xb_2 - \Delta, \xb_2 + \Delta]$. The probability for these two measured results is
\beq
p (\xb_1, t_1, \xb_2, t_2 ) = \langle \psi | g^\dag (t_1,t_0) P_{\xb_1} g^\dag (t_2,t_1) P_{\xb_2} g(t_2,t_1) P_{\xb_1} g (t_1,t_0)| \psi \rangle
\label{probx}
\eeq
where
\beq
P_{\xb} = \int_{\xb-\Delta}^{\xb+\Delta} dx | x \rangle \langle x |
\eeq
is a projector onto the range $ [\xb-\Delta, \xb+\Delta]$ and $g(t_1,t_0)$ denotes the propagator in the region $x>0$.
The precise form of the propagator depends on the boundary conditions on the states imposed at $x=0$. There is a one-parameter family that leads to a self-adjoint Hamiltonian, of the form
\beq
\psi (0) + i  \alpha \psi'(0) = 0
\eeq
where $\alpha$ is real \cite{Mom}.
For simplicity, we focus on the simplest case $\alpha = 0$ in which the states are required to vanish at $x=0$ and in this case the propagator in $x>0$  is given by the simple method of images expression
\bea
g(x_1,t_1|x_0,t_0) &=& \theta (x_1) \theta (x_0)
\left( \frac { m } { 2 \pi \hbar (t_1 - t_0) } \right)^\half
\nonumber \\
& \times &
\left[ \exp \left( \frac { i m  ( x_1 - x_0)^2 } { 2 \hbar (t_1 - t_0) } \right)
- \exp \left( \frac { i m  ( x_1 + x_0)^2 } { 2 \hbar (t_1 - t_0) } \right)
\right].
\label{gres}
\eea
Semiclassically, the first term clearly corresponds to the direct classical path between the initial and final spacetime points and the second term corresponds to a reflected path (or equivalently, to the path coming from the image point $-x_0$). Other choices of boundary conditions yield different propagators but all have the same feature of including both direct and reflected paths.

To determine the momentum from the probability Eq.(\ref{probx}) we consider the conditional probability of finding the particle near $\xb_2$ at time $t_2$, given that it was near $\xb_1$ at time $t_1$, which is given by
\beq
p( \xb_2, t_2 | \xb_1, t_1 ) = \frac { p (\xb_1, t_1 , \xb_2, t_2 ) } { p ( \xb_1, t_1 ) }.
\label{probx2}
\eeq
By comparing with earlier cases for the whole real line \cite{ToF}, it is reasonably easy to see what happens without doing any further calculation. This probability distribution will have two peaks, one at the expected value for an incoming state with momentum $p_0$, namely $ \xb_2 = \xb_1 + (p_0/m)(t_2 - t_1) $. The other peak is
at $ \xb_2 =  \xb_1 - (p_0/m) (t_2 - t_1) $ corresponding to the parts of the incoming state that already reached the origin prior to $t_1$ and been reflected back.
Hence the measured momentum probability will be concentrated about both $p_0$ and $-p_0$ without a possibility of distinguishing between the two. The magnitude of the momentum is measured unambiguously but the sign of the momentum is intrinsically ambiguous, due to reflection.

Of course the case described above is the extreme case of an incoming plane wave with no spatial localization so the state arrives at the first measurement at a very wide spread of times.
The opposite extreme is a wave packet tightly peaked in both position and momentum which arrives at the first measurement at a precise time and the probability Eq.(\ref{probx}) is strongly peaked about the direct momentum $p_0$ with negligble peak about the reflected one. But for a more general state, there always will be some peaking about the reflected momentum and hence ambiguity in the momentum measurement. This is the physically intuitive reason underlying the imprecision in the sign of the momentum operator on the half-line associated with its lack of self-adjointness.

This intuitive link between reflection and lack of self-adjointness is further substantiated
in another closely related but different case, namely the momentum operator in a finite spatial interval $[a,b]$. In this case the momentum operator is in fact self-adjoint, for suitable periodic boundary conditions on the state, the simplest of which is
\beq
\psi (a) =  \psi (b).
\eeq
(see for example Ref.\cite{Mom}).
One can again ask how this reconciles with measurement of momentum in this interval, using sequential position measurements -- is there any imprecision due to reflection effects? The point here is that the boundary conditions mean the configuration space has the topology of the cylinder and the semiclassical interpretation of the propagator is that it consists of straight line paths on the cylinder. There is no reflection at the boundaries, as there is in the half-line case. This is consistent with self-adjointness.

Given these features of the momentum on the half-line,
we may now draw a direct connection with the Aharonov-Bohm operator. In exploring the quantization of the classical arrival time $ - m x /p$, it is not at all obvious at the classical level why this object leads to an operator which is not self-adjoint. To explore this,
one might take as a simple starting point the simple expression $\theta ( t + mx/p ) $, which may be used to address the simple question of whether the arrival time is greater or less than a given time $t$. The arrival time is then constructed using
\beq
- \frac { mx}{p} = \int_{-\infty}^{\infty} dt \ t \ \frac {d} {dt} \theta \left( t + \frac {mx}{p} \right)
\eeq
Quantizing $\theta ( t + mx/p ) $ directly is problematic since $ - m x/p$ does not turn into a self-adjoint operator so it is natural to consider alternative but equivalent classical expressions obtained by scaling out $p$ to relate it to the expression $ \theta ( x + pt/m)$, which clearly can be turned into a self-adjoint operator. (The latter expression is also a natural starting point for measurement inspired models as we shall see.) However, the sign of $p$ is important in this scaling and we in fact obtain
\beq
\theta \left( t + \frac {mx}{p} \right) = \theta (p) \theta \left( x + \frac {pt} {m} \right)
+ \theta (-p) \theta \left( -x - \frac {pt} {m} \right)
\eeq
This is just $\theta (x) \theta (p) + \theta (-x) \theta (-p)$, shifted along the classical equations of motion, precisely the object that cannot be made self-adjoint as  we have seen. For example,
with a particular choice of operator ordering the quantization of the right-hand side is
\beq
\exp \left( \ih H t \right)
\left[ \theta (  \hat x ) \theta (\hat p ) \theta ( \hat x )
- \theta (  -\hat x ) \theta (-\hat p ) \theta ( -\hat x ) \right]
\exp \left( -\ih H t \right)
\eeq
A different operator ordering would lead to
the expressions of the form $\theta (\hat p) \theta (\hat x) \theta (\hat p)$ which are also not self-adjoint, as noted already.
Hence, the quantization of the classical arrival time expression $ - m x/p$ boils down to the expression $ \theta (\hat x ) \theta (\hat p ) \theta ( \hat x ) $, precisely the underlying non-self-adjoint object encountered in the momentum operator on the half line. The same remarks concerning reflection therefore apply.

By contrast, for the classical expression $ - m x / | p|$, similar steps yield
\beq
\theta \left( t + \frac { m x } { |p| } \right) = \theta \left( x + \frac { | p | t } { m} \right)
\eeq
The right-hand side, when quantized, is simply $\theta (\hat x)$ unitarily evolved in time with the pseudo-Hamiltonian $ \hat \xi = \hat p | \hat p | / 2 m$, so is self-adjoint.

In summary, the key difference between the Aharonov-Bohm operator and the three self-adjoint variants of it is that the Aharonov-Bohm operator involves specification of whether the states are incoming or outgoing. This involves specifying the signs of the momenta on the two half-lines $x>0$ and $x<0$.
Restriction to the half-line in quantum mechanics creates quantum-mechanical reflection which renders the sign of the momentum imprecise. This is the physical reason why we would expect that the Aharonov-Bohm operator is not self-adjoint.

\section{Difficulties with the Standard Arrival Time Operators}

Aside from the self-adjointness issue, all four of the operators discussed in Section II suffer from a number of problems. Firstly, their relation to actual measurements, or at least to simple measurement models is not obvious since there is no physical system that couples to the Aharonov-Bohm operator or any of the above variants of it.
Secondly all the above arrival time operators yield an arrival time distribution which is the Kijowski distribution or variants thereof for all ranges of momenta.
Simple measurement models (such as those based on a complex potential \cite{Meas,kin1} or stopwatch \cite{kin3,Oppenheim}) agree with this for large momenta. But crucially, for small momenta, and more specifically for arrival time measurements more precise than $ \hbar m/p^2$ (the energy time \cite{HaYe,AOPRU}), reflection off the detector becomes significant and measurement models typically yield a distribution proportional to the kinetic energy density
\beq
\Pi (\tau) =
N \langle \psi_\tau | \hat{p} \delta(\hat{x}) \hat{p} | \psi_\tau \rangle,
\label{lowp}
\eeq
where $N$ is a model-dependent normalization factor \cite{Meas,kin1,kin3,Oppenheim,AnSa}.
This behaviour does not arise in any of the standard arrival time operators at small momentum.

Thirdly, even after normalizable states are constructed from the eigenstates of $\hat T_{AB}$
and its three modifications (by superposing over narrow ranges of $\tau$, as described in Ref.\cite{Oppenheim}), they all go like $p^\half$ for small $p$, which means that they have infinite $(\Delta x)^2$, so have poor spatial localization properties, contrary to the intuitive notion of what an arrival time state should look like. The last two problems are issues around small momentum behaviour and could be solved by an arrival time operator whose eigenstates go like $p$ for small $p$. The main result of the remainder of this paper, and the solution to these three problems, is the construction of an arrival time operator directly inspired by simple measurement models.

We briefly mention the work of Grot, Rovelli and Tate, who removed the singularity at $p=0$ in $ \hat T_{AB}$ by a somewhat artificial regularization procedure \cite{GRT}. This produced a self-adjoint operator but the low momentum behaviour produces a spatial spread even more severe than the examples above \cite{Oppenheim}, reiterating the need for a physically motivated handling of the low momentum regime. See also Ref.\cite{MiTo} for other criticisms of traditional arrival time operators.

\section{Measurement Models}


To motivate our proposed new arrival time operator and also to substantiate the arrival time distribution in the low momentum regime Eq.(\ref{lowp}) we consider two measurement models for arrival time.

A simple measurement model for the probability of a particle crossing the origin in a given time interval $[0,\tau]$ involves spatial measurements onto $x>0$ and $x<0$ described by projectors $P = \theta(\hat{x})$
and $\bar{P} = \theta(-\hat{x})$, and we simply check to see if the particle is on opposite sides of the origin at the initial and final times. The probability for crossing is then
\beq
\label{E:trace}
p(0, \tau) = \langle \psi| \bar{P}P(\tau) \bar{P} | \psi \rangle + \langle \psi| P \bar{P}(\tau) P | \psi \rangle.
\eeq
Since $ d P (t)/dt =  \hat J (t)$, where $\hat J(t)$ is the current operator Eq.(\ref{cur}), this may be rewritten
\beq
p(0, \tau) = \int_{0}^{\tau} dt  \langle \psi | \bar{P} \hat{J}(t) \bar{P} | \psi \rangle
-\int_{0}^{\tau} dt  \langle \psi | {P} \hat{J}(t) {P} | \psi \rangle,
\label{cross}
\eeq
thus indicating the appearance of the current operator in arrival time probabilities derived from a measurement model (as it does in models, e.g. Ref.\cite{kin3}). This is not in fact the Kijowski distribution (integrated over a time range) Eq.(\ref{Kij}). However, estimates involving gaussian states show that it is very close either for states strongly peaked in momentum, or for time intervals large compared to the energy time, $\hbar m/p^2 $.

We may use Eq.(\ref{cross}) to see the origin of the low momentum regime formula Eq.(\ref{lowp}) as follows. The formula Eq.(\ref{lowp}) is the appropriate one in the case of very frequent measurement in which there is significant reflection off the detector. To model this,
we suppose that prior to $t=0$, the system has been subjected to frequent projections onto $x<0$, so that the state at $t=0$ has the form
\beq
|\psi  \rangle = \bar P (t_n) \cdots \bar P(t_2)  \bar P( t_1) | \phi_0 \rangle
\label{PP}
\eeq
for a sequence of times $ 0 > t_n > \cdots > t_2 > t_1 $ and for some initial state $| \phi_0 \rangle $ in the distant past.  If the interval between these projections is sufficiently small (in comparsion to the energy time $ \hbar m  /p^2 $) the evolution in Eq.(\ref{PP}) will in fact be well-approximated by the restricted propagator for the half-line $x<0$, given by Eq.(\ref{gres}) adapted to the case $x<0$, so there will be reflection off the origin \cite{HaYe}.
At $t=0$ the state will therefore have the approximate form $\psi(x) = \theta (-x) ( \phi (x) - \phi(-x))$, for some $\phi(x)$,
so $\psi(x)$ is zero at $x=0$ but has non-zero derivative. The arrival time distribution is then
\bea
\Pi (\tau) &=& \langle \psi | \hat J(\tau) | \psi \rangle
\nonumber \\
&=& - \frac{i\hbar}{2m}[\psi^*(0,\tau)\psi'(0,\tau) - \psi(0,\tau)\psi'^*(0,\tau)],
\eea
where the dash denotes spatial derivative.
This is zero at $\tau=0$ so we have to expand the wave function for small $\tau$.
We make use of the free particle propagator and write
\beq
\psi(0, \tau) = \int_{-\infty}^{0} dy \left(\frac{m}{2\pi i \hbar \tau}\right)^\frac{1}{2} e^{\frac{imy^2}{2\tau\hbar}}\psi(y, 0),
\eeq
where we have used the fact that $\psi(x > 0, 0) = 0$. Making the change of variables $y =z \tau^\frac{1}{2}$,
expanding $\psi( z \tau^\half , 0)$, and recalling that $\psi(0,0) = 0$, we thus obtain
\beq
\psi(0, \tau)  \approx  \left(\frac{m\tau}{2\pi i\hbar}\right)^\frac{1}{2}\psi '(0,0)  \int_{-\infty}^0 dz\, z\, e^{\frac{imz^2}{2\hbar}}.
\eeq
Evaluating the integral we then find
\beq \label{E:meas}
\langle J(\tau) \rangle \approx \frac{1}{2 \pi^\half} \left(\frac{\hbar}{m}\right)^{\frac{3}{2}}\tau^\half \ |\psi'(0,0)|^{2}.
\eeq
This is of the desired form, Eq.(\ref{lowp}) since
\beq
\langle \psi | \hat p \delta ( \hat x ) \hat p | \psi \rangle = \hbar^2 \left| \psi'(0,0) \right|^2
\eeq
(A similar derivation was given in Ref.\cite{Sokolowski} with different aims).
Hence the low momentum regime result arises simply because the very frequent measurement causes the wave function to vanish at the origin so it is necessary to expand the
average current around this for small times.

A second simple model for measuring the arrival time, considered by numerous authors \cite{Oppenheim,kin3}, consists of a stopwatch -- a system with coordinate $y$ and zero Hamiltonian which couples to the particle through the interaction $ p_y \theta ( - x) $. Classically, for a particle approaching from the left it therefore causes a shift in the stopwatch variable $y$ for the entire time the particle is in $x<0$, stopping when the particle reaches the origin, with final value
\beq
y (T) -y(0) = \int_0^T dt \ \theta \left( - x - \frac {p}{m} t \right),
\label{stopw}
\eeq
where $T$ is taken to be very large. This is easily seen to be equal classically to $ - m x / p $ for $p>0$. In the quantum-mechanical analysis of this system we can see the form of the interaction between particle and stopwatch via the $S$-matrix,
\beq
S ={\mathbb T} \exp \left( - \ih \lambda \int_0^T dt \ \hat p_y (t) \theta \left( - \hat x - \frac{ \hat p} {m} t \right) \right)
\label{Smat}
\eeq
where ${\mathbb T}$ is the usual time ordering operator and $\lambda$ is the coupling constant. The stopwatch has zero free Hamiltonian so $ \hat p_y (t) $ is independent of $t$.
For weak couplings we therefore see that the combination of the particle variables that the stopwatch ``sees" is precisely the quantization of the right-hand side of Eq.(\ref{stopw}). It does not couple to anything like the Aharonov-Bohm operator.

It is also of interest to rewrite the right-hand side of Eq.(\ref{stopw}) to indicate its connection to the current. We have,
\beq
\int_0^T dt \ \theta \left( - x - \frac {p}{m} t \right)
= \int_0^T dt \ t J(t)
\label{stopw2}
\eeq
where $J(t) = (p/m) \delta (x + pt/m)$ is the classical current and we have dropped a boundary term since $ \theta ( - x - pt/m) $ is zero for large $t$ with positive $p$. We could also contemplate a similar integration by parts for the corresponding quantum operators but there it is less obvious that the boundary term may be dropped. Nevertheless, both sides of Eq.(\ref{stopw2}) give two alternative starting points for the construction of arrival time operators, classically equivalent to $ -m x / p$, but potentially different in the quantum case and arguably more relevant to measurement models. The expression on the left-hand side is difficult to handle as a quantum operator so we focus on a version of the right-hand side in what follows.



\section{New Operator and its Properties}

Motivated by the above observations and in particular by the key role the current plays in measurement models,
we look for a new time operator defined in terms of the current operator $\hat{J}(t)$, Eq.(\ref{cur}). We begin by noting the classical result indicated already, namely
\beq \label{classical3}
-\frac{mx}{|p|}= \int_{-\infty}^{\infty} dt\, t\, J(t),
\eeq
where $J(t) $ is the classical current, here valid for all values of $x$ and $p$.
Quantization of the left-hand side yields the KDM operator Eq.(\ref{DMop}). Here, however, we
instead take the right-hand side as the starting point for quantization and this leads to a different operator, namely
\beq \label{current}
\hat{T} = \int_{-\infty}^{\infty} dt\, t\, \hat{J}(t),
\eeq
whose properties we study. To be clear, there are two motivations for examining this expression. Firstly, it has a potentially closer connection to measurements than the KDM operator, as indicated. But secondly, and independently of this, it may be simply regarded as an exploration of the consequences in the quantum theory of taking a different but classically equivalent starting point.

To evaluate the integral in Eq.(\ref{current}) we first note the operator identity
\beq
\frac { \hat p t } { m} \ \delta \left( \hat x + \frac { \hat p t } { m} \right)
= - \hat x \ \delta \left( \hat x + \frac { \hat p t } { m} \right)
\eeq
The remaining time integral may be evaluated
by sandwiching between momentum states. We have
\bea
\int_{-\infty}^{\infty} dt
\ \langle p_1 | \delta \left( \hat x + \frac { \hat p t } { m} \right) | p_2 \rangle
&=&
\langle p_1 | \delta ( \hat x ) | p_2 \rangle \int_{-\infty}^{\infty} dt
\ \exp \left( \frac { it } {2 m \hbar } (p_1^2 -p_2^2) \right)
\nonumber \\
&=& 2m \ \delta (p_1^2 - p_2^2 )
\nonumber \\
&=& \frac {m} {|p_1|} \ \left( \delta (p_1 - p_2) + \delta (p_1 + p_2) \right)
\nonumber \\
&=& \langle p_1 | \frac { (1+\hat R) } { | \hat p| } | p_2 \rangle
\label{tint}
\eea
where we have introduced the reflection operator $\hat R$, defined by $ \hat R | p \rangle
= | - p \rangle$. We thus find that our time operator may now be written
\beq
\hat{T} =  -\frac{m}{2}\left(\hat{x}\frac{1}{|\hat{p}|}(1+\hat{R}) + \frac{1}{|\hat{p}|}(1+\hat{R})\hat{x}\right),
\label{newop}
\eeq
It may be written in terms of the KDM operator Eq.(\ref{DMop}) as
\beq
\hat{T} = \hat{T}_{KDM} +  \frac { i \hbar m } { 2 \hat p | \hat p |} \ \hat R
\label{newop2}
\eeq
Like the KDM operator, it is self-adjoint, as we shall confirm.
The extra term, involving reflection, arises because of the time integral in Eq.(\ref{tint}), which is precisely how reflection effects arise in scattering theory in the usual $S$-matrix expansion

We also note  that our operator bears a close comparison with the dwell time operator. This is the operator describing the time spent by a particle in a spatial region $[0,L]$, defined by
\beq \label{dwellop}
\hat{T}_{D} = \int_{-\infty}^{\infty} dt \ e^{ \ih H t} \ P_L e^{ -\ih H t},
\eeq
where $H = \hat p^2 / 2m $ and $P_L  = \int_{0}^{L} dx | x \rangle \langle x | $ is a projector on the region $[0,L]$ \cite{dwell}.
The dwell time operator commutes with the Hamiltonian so is unaffected by the Pauli theorem and indeed it is self-adjoint. We would expect self-adjointness on physical grounds. In a measurement model for dwell time (see for example Ref.\cite{kin3}), one might consider a clock model in which the Hamiltonian (which we take to be self-adjoint) includes a coupling between the clock and the operator $ P_L $ and the clock variables would then ``see" the expression Eq.(\ref{dwellop}) in an $S$-matrix expression, analogous to Eq.(\ref{Smat}).

The dwell time operator may also be written more explicitly as
\beq \label{dwell2}
\hat{T}_{D}=\frac{mL}{|\hat{p}|}\left(1 + e^{- \ih \hat p L }\frac{\sin{(\hat{p}L/\hbar)}}{(\hat{p}L/\hbar)}\hat{R} \right),
\eeq
in which one can clearly see reflection effects.
Classically, the dwell time can be written as the difference between two arrival times, at $x=L$ and $x=0$
\beq
\frac {mL} {|p|} =  - \frac { m ( x-L)} { |p|} + \frac { m x } { |p|}
\label{classD}
\eeq
but not in general in the quantum case Eq.(\ref{dwell2}). However, it is true both for large momenta, $ pl \gg \hbar$, where the quantum case essentially coincides with the classical one as one might expect, but more interestingly also for low momenta, $ p L \ll \hbar $, where one can see that our new arrival time operator satisfies
\bea
\hat T_D & \approx & \frac{mL} { | \hat p | } \left( 1 + \hat R \right)
\nonumber \\
&=& e^{ \ih L \hat p} \hat T  e^{ -\ih L \hat p } - \hat T.
\eea
Hence the reflection term occurring in our new arrival time operator is closely related to a similar term appearing in the more familiar dwell time operator.

Returning to our new time operator, we consider its covariance properties.
Note that it has the same commutation relation with the Hamiltonian as the KDM operator, namely
\beq
[ H, \hat T ] = i \hbar \ \epsilon (\hat p )
\eeq
since the extra term in Eq.(\ref{newop2}) commutes with $H$. Applying the unitary Hamiltonian evolution operator to the eigenvalue equation,
\beq
\hat T | \phi_{\tau} \rangle = \tau | \phi_{\tau} \rangle
\eeq
we find that
\beq
\hat T \left( e^{ - \ih H t} | \phi_{\tau} \rangle \right) = ( \tau + \epsilon (\hat p) t )
\left( e^{ - \ih H t} | \phi_{\tau} \rangle \right).
\eeq
For the KDM case, we can solve this equation for $ e^{ - \ih H t} | \phi_{\tau} \rangle $ by splitting into positive and negative momentum states, thereby finding that
the positive momentum states are shifted forwards in time and the negative momentum states backwards (as one can see also in the explicit eigenvalues, Eq.(\ref{dmstates})), so there is covariance in the positive and momentum sectors separately. However, for our new operator $\hat T$, the presence of the reflection term mixes up the positive and negative momentum sectors and it can be shown that the spectrum does not transform in any simple way under unitary time shifts.  This difference between $\hat T$ and the KDM operator may also be seen by considering their commutation with the pseudo-energy $\hat \xi = \hat p | \hat p | / 2 m $. We have
\beq
[ \hat \xi, \hat{T}_{KDM} ] = i \hbar
\eeq
which implies the simple shift properties stated above. However,
the commutator with $\hat T$ has an extra term,
\beq
[ \hat \xi, \hat T ] = i \hbar \left( 1 + \half \hat R \right)
\eeq
from which no simple covariance properties follow, as one can show.
These results mean that we do not expect any obvious covariance properties in the spectrum of $\hat T$.
However, since the extra term is clearly quantum-mechanical in nature, we would expect covariance behaviour similar to the KDM case in the semiclassical regime.

\section{Spectrum and Arrival Time Probabilities}

Consider now the spectrum of the arrival time operator Eq.(\ref{newop}).
The eigenvalue equation in momentum space is
\beq \label{E:eigen}
- i \frac{m\hbar}{2} \left[\frac{\partial}{\partial p} \frac{1}{|p|}\left(1 + \hat{R}\right) + \frac{1}{|p|} \left(1 + \hat{R}\right)\frac{\partial}{\partial p} \right] \phi_{\tau}(p) = \tau \phi_{\tau}(p).
\eeq
The eigenstates can be written as the sum of their symmetric and anti-symmetric parts
$\phi_{\tau}(p) = \phi_{\tau}^S(p) + \phi_{\tau}^A(p)$
and the eigenvalue equation reduces to a coupled system of first order equations
\beq
\frac{\partial}{\partial p} \left( \frac{1}{|p|} \phi_{\tau}^S(p) \right) = \frac{i \tau}{m\hbar} \phi_{\tau}^A(p),
\eeq
and
\beq \label{sym}
\frac{\partial}{\partial p} \phi_{\tau}^A(p) =  \frac{i \tau |p|}{m \hbar} \phi_{\tau}^S(p).
\eeq
Solving for the antisymmetric part of the state we obtain the second order differential equation
\beq
\frac{\partial^{2}}{\partial p^{2}}\phi_{\tau}^{A}(p)-\frac{2}{p}\frac{\partial}{\partial p}\phi_{\tau}^{A}(p) +\left(\frac{\tau}{m\hbar}\right)^{2}p^{2}\phi_{\tau}^{A}(p)=0.
\label{2ODE}
\eeq
It may be shown that this equation has two linearly independent solutions in terms of Bessel functions, an antisymmetric one,
$ p |p|^\half J_{\frac{3}{4}}\left(\frac{p^{2}\tau}{2m\hbar}\right)$
and a symmetric one $|p|^{\frac{3}{2}} J_{-\frac{3}{4}}\left(\frac{p^{2}\tau}{2m\hbar}\right)$ which is irrelevant so is dropped.  Inserting this antisymmetric solution into Eq.(\ref{sym}) and using properties of the Bessel functions \cite{Bess} we obtain
$ \phi_{\tau}^{S}(p)$ and the full normalized solution is then found to be
\beq \label{E:states}
\phi_{\tau}(p) = \frac{\tau^{\frac{1}{2}}}{\sqrt{8}m\hbar}\left(
|p|^{\frac{3}{2}}J_{-\frac{1}{4}}\left(\frac{p^2 \tau}{2m \hbar}\right)
+ i p |p|^{\frac{1}{2}}J_{\frac{3}{4}}\left(\frac{p^2 \tau}{2m \hbar}\right)
\right).
\eeq
(Here, a possible overall factor of $i$ is fixed by noting that the solution must satisfy
$\phi^*(p) = \phi(-p)$).
These eigenstates may be shown, at some length, to be orthonormal and complete and thus $\hat{T}$ is a self-adjoint operator.
Using the asymptotic forms of the Bessel functions, we obtain approximations for the eigenstates in the large and low momentum regimes respectively,
\beq
\phi_{\tau}(p) \sim \begin{cases}  \ e^{ -  \frac{\pi i}{8}\epsilon(p) }\left(\frac{|p|}{2 \pi m \hbar}\right)^{\half} \exp \left( i \epsilon (p) \frac{p^2\tau}{2m\hbar} \right) &  \frac{p^2 \tau}{2m\hbar} \gg 1\\\
\frac{1}{2\Gamma\left(\frac{3}{4}\right)}
\frac { \tau^{\frac{1}{4}}} {\left(m \hbar\right)^{\frac{3}{4}}}   |p| & \frac{p^2 \tau}{2m\hbar} \ll 1.
\end{cases}
\label{asymp}
\eeq
As anticipated, the eigenstates do not shift in a simple way under time translations, except in the large momentum regime, where they have the same covariance properties as the KDM states.

The arrival time probability following from our new operator for a time interval $[\tau_1, \tau_2]$ is now given by
\beq
p (\tau_1,  \tau_2) = \int_{\tau_1}^{ \tau_2} d \tau  \left| \langle \psi | \phi_\tau \rangle \right|^2.
\label{prob12}
\eeq
We compare this directly with the measurement model, Eq.(\ref{cross}), so we set $\tau_1 = 0$ and consider the interval $[0,\tau_2]$.
If $\tau_2 $ is large compared to the energy time $ m \hbar /p^2 $, the time integral in Eq.(\ref{prob12}) takes contributions from both large and small momentum regimes in Eq.(\ref{asymp}). It is easy to verify that the large momentum regime will give the dominant contribution.
The eigenstates in that regime are of the form (\ref{dmstates}) and thus the probability distribution is similar to that for $\hat{T}_{KDM}$, i.e. a variant of the Kijowski distribution \cite{DMdis}.
Note, however, there is a difference in that there is a phase depending on $\epsilon(p)$ which implies that the terms in the distribution representing interference between positive and negative momenta will have a different phase to that in the distribution derived from $\hat T_{KDM}$.
This difference is not surprising given the presence of a reflection term in our operator.
We thus get broad agreement with Eq.(\ref{cross}).

If the time $\tau_2$ is short compared to the energy time in Eq.(\ref{prob12}), the low momentum regime applies.
The eigenstates are of the form $ \phi_{\tau}(p) \approx C  |p| $,
where $C$ is read off from Eq.(\ref{asymp}) and we easily obtain
\beq
|\langle \psi|\phi_\tau \rangle |^2 \approx \frac{\pi}{2 \left(\Gamma\left(\frac{3}{4}\right)\right)^2}
\frac { |\tau|^\half } {m^{ \frac{3}{2} } \hbar^\half }
\langle \psi | |\hat p| \delta ( \hat x ) |\hat p| | \psi \rangle,
\label{KED2}
\eeq
which is the kinetic energy density, Eqs.(\ref{lowp}), (\ref{E:meas}), the physically expected result in this regime. The numerical prefactors are not exactly the same as in Eq.(\ref{E:meas}) -- they differ by about 20 percent
but the overall factors in Eq.(\ref{KED2}) are fixed by the completeness relation of the eigenstates $\phi_\tau (p)$. There is no reason to expect perfect agreement since the two formulae have different origins. So again we get broad agreement with the measurement model.

Hence our new arrival time operator Eq.(\ref{newop}) gives the physically expected results for both large and small momentum regimes, and in particular, gives the behaviour anticipated by measurement models in the low momentum regime, behaviour not captured by any of the standard arrival time operators described in Section II.

However, it is important to note that approximate agreement of the probability Eq.(\ref{prob12}) with the low momentum arrival time formula Eq.(\ref{lowp}) is achieved only for an interval $[0,\tau_2]$ with $\tau_2$ small. This is all that is required since we are comparing with the formula derived from measurement, Eq.(\ref{cross}), which concerns a time interval close to zero. The initial time $\tau=0$ has a distinguished status in this model since the state is prepared then (and in the stopwatch model it is the initial time at which stopwatch and particle are in a factored state).
Hence this feature is physically expected. It is also mathematically consistent since, as noted, the spectrum of our operator does not have any obvious covariance properties under time translation so there is no inconsistency in the model possessing a distinguished point in time.
For more general time intervals $[\tau_1, \tau_2]$ we get the low momentum result
only for time intervals where {\it both} end-points are close to $\tau = 0$.
For times $\tau_1, \tau_2$ which are very close, but large, the probability Eq.(\ref{prob12}) is still an integral over a range of time of a distribution similar to the Kijowski distribution, not the low momentum formula.

These features mean that the regimes in this model are not in general defined by the size of the interval $ | \tau_2 - \tau_1 | $ compared to the energy time, as one might have thought from some measurement models, except for the special case of the interval $ [0,\tau_2]$.
To obtain a result of this form would require a more elaborate operator containing an explicit parameter modeling the precision of the measurement. This is of interest to construct and will be pursued elsewhere.

Furthermore, as indicated earlier, there are two different motivations for looking at the new time operator and the second motivation is the different question of exploring the consequences of taking an alternative but equivalent classical starting point, without necessarily making any claims about a connection to measurement models.
In that case, our new operator shows that a different classical starting point leads to a probability distribution $ | \langle \psi | \phi_\tau \rangle |^2 $
which is a variant of the Kijowski distribution for large momenta. At small momenta, the alternative starting point yields improved behaviour in the eigenstates, since the states go like $|p|$ instead of $p^\half$. This means that arrival time states obtained by superposing small ranges of $\tau$ have finite $(\Delta x)^2 $, and hence better localization properties than the four standard results discussed in Section II. So interpreted purely as an alternative classical starting point our new operator gives useful and physically reasonable results.

\section{Summary and Conclusion}

In the construction of arrival time operators, different classical starting points lead to inequivalent quantum operators with different physical predictions. The Aharonov-Bohm operator and its variants do not capture the expected physical behaviour in the low momentum regime, suggesting that a new starting point is called for.
In this paper we constructed an arrival time operator taking a different starting point based on the current, inspired by measurement models and classically equivalent to $ - m x / |  p | $, thus obtaining an arrival time probability giving the expected physical behaviour in both large and small momentum regimes. In addition, we have shed some light on the intuitve origins of self-adjointness or its absence in both arrival times and the momentum operator on the half-line -- it relates to the imprecision arising due to reflection when $x$ is restricted to a half-infinite range.

\section{Acknowledgements}

We are grateful to Gonzalo Muga for useful conversations over a long period of time.
This work was supported in part by EPSRC Grant No. EP/J008060/1.

\end{document}